\documentclass[prl,twocolumn,floatfix,superscriptaddress]{revtex4}

\usepackage{graphicx,amssymb,amsmath}
\usepackage{color,soul,bbold}

\begin{document}

\title{Coupling of Plasmon Modes in Graphene Microstructures}

\author{Parinita Nene}
\author{Jared H. Strait}
\author{Wei-Min Chan}
\author{Christina Manolatou}
\affiliation{School of Electrical and Computer Engineering, Cornell University, Ithaca, NY, 14853}
%\author{Joshua W. Kevek}
%\affiliation{Laboratory of Atomic and Solid State Physics and the Kavli Institute at Cornell for Nanoscale Science, Cornell University, Ithaca, NY 14853}
\author{Sandip Tiwari}
\affiliation{School of Electrical and Computer Engineering, Cornell University, Ithaca, NY, 14853}
\author{Paul L. McEuen}
\affiliation{Laboratory of Atomic and Solid State Physics and the Kavli Institute at Cornell for Nanoscale Science, Cornell University, Ithaca, NY 14853}
\author{Farhan Rana}
\email{fr37@cornell.edu}
\affiliation{School of Electrical and Computer Engineering, Cornell University, Ithaca, NY, 14853}

\begin{abstract} 
A variety of different graphene plasmonic structures and devices have been proposed and demonstrated experimentally. Plasmon modes in graphene microstructures interact strongly via the depolarization fields. An accurate quantitative description of the coupling between plasmon modes is required for designing and understanding complex plasmonic devices. Drawing inspiration from microphotonics, we present a coupled-mode theory for graphene plasmonics in which the plasmon eigenmodes of a coupled system are expressed in terms of the plasmon eigenmodes of its uncoupled sub-systems. The coupled-mode theory enables accurate computation of the coupling between the plasmon modes and of the resulting dynamics. We compare theory with experiments performed on the plasmon modes in coupled arrays of graphene strips. In experiments, we tune the coupling by changing the spacing between the graphene strips in the array. Our results show that the coupling parameters obtained from the coupled-mode theory and the plasmon frequency changes resulting from this coupling agree very well with experiments. The work presented here provides a framework for designing and understanding coupled graphene plasmonic structures.
\end{abstract}

\maketitle
Graphene, a single layer of carbon atoms arranged in honeycomb lattice, has emerged as an important material for plasmonics~\cite{Ju11,HYan12,Rana08,DasSarma09,Hwang07,Xia12,Koppens11,Low14}. The high carrier mobility and the widely tunable conductivity in graphene together with the ability to fabricate graphene microstructures of different sizes implies that plasmons in graphene structures can have high quality factors and frequencies tunable from a few THz to more than 100 THz. Graphene plasmonic structures have the potential to form building blocks for novel THz/IR devices such as detectors, emitters, oscillators, switches, filters, and sensors. Several theoretical works have explored techniques to compute the modes in individual as well as in arrays of graphene plasmon resonators~\cite{Peres13,Fallahi12,Thongrattanasiri12,Hanson08,Ferreira12,Strait13}. In order to realize the full potential of graphene plasmonics, and develop the ability to combine several graphene plasmonic resonators and engineer complex device structures, suitable techniques are needed model the interactions between plasmonic resonators in simple, yet effective and accurate, ways. In the field of microphotonics, the equivalent role is played by coupled-mode theories\cite{Haus91,Yariv73}. In coupled-mode theories, the field of a coupled system is expanded in terms of the fields of the eigenmodes of its uncoupled sub-systems\cite{Haus91,Yariv73}. Accurate computation of the coupling parameters and the response of the coupled system without detailed first-principles electromagnetic simulations are few of the main benefits of coupled-mode theories. Coupled-mode theories have proven to be extremely effective tools in designing and understanding complex optical integrated structures\cite{Manolatou01}. In the field of graphene plasmonics, complex devices incorporating several coupled plasmonic resonators have been proposed and demonstrated for various applications~\cite{Shvets13,Shi13,Amin13,Fang14,Ju11,HYan12,Sensale12}. In this paper, we present a coupled-mode theory for graphene plasmonic structures. We show that the coupling parameters that describe the interaction between the plasmon modes of different resonators can be calculated accurately from the plasmon eigenmodes. We also present experimental results for the interactions among graphene plasmon modes in coupled plasmonic strip structures, in which the couplings are tuned by varying the spacing between the strips, and compare the measurements with the theory in a quantitative way. We show that the results from the coupled-mode theory agree well with the experimental data as well as with electrodynamic simulations using finite-difference time-domain (FDTD) technique.  

\begin{figure}[tbp]
	\centering
		\includegraphics[width=.45\textwidth]{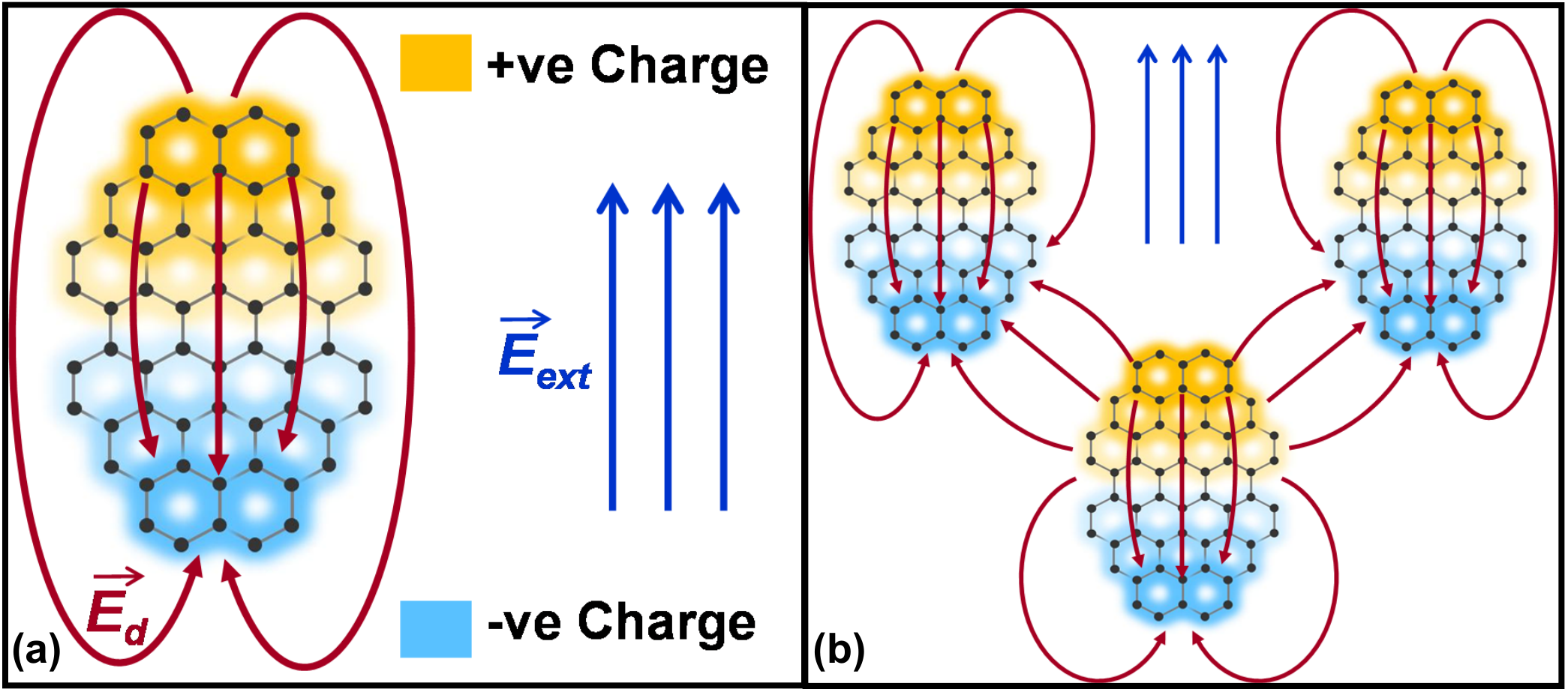}
	\caption{(a) The  plasmon mode of a single resonator is depicted along with the depolarization field $\vec{E}_{d}$ and the driving field $\vec{E}_{ext}$. (b) The coupled plasmon mode of three resonators is shown. The coupling of the resonators results in the depolarization field of one resonator exciting the plasmon mode of another resonator.}
  	\label{fig:CMT}
\end{figure}
Consider a system of coupled plasmonic resonators, as shown in Fig.\ref{fig:CMT}. Suppose the current density of the $\alpha$-th plasmon eigenmode of the $n$-th resonator is $\vec{K}^{\alpha}_n$. Previously, the authors have shown that if retardation effects are ignored then $\vec{K}^{\alpha}_n$ satisfies the eigenvalue equation~\cite{Strait13}.  
\begin{equation}
\frac{\sigma(\vec{r})}{4\pi\epsilon_{avg}\tau}  \int \textrm{d}^{2}\vec{r}' \bar{\bar{f}}(\vec{r}-\vec{r}') \cdot \vec{K}^\alpha_n(\vec{r}')  = (\omega^\alpha_n)^{2} \vec{K}^\alpha_n(\vec{r}) \label{eq:K0}
\end{equation}
The tensor $\bar{\bar{f}}(\vec{r}-\vec{r}')$ equals $\vec{\nabla}_{\vec{r}} \vec{\nabla}_{\vec{r}'} |\vec{r}-\vec{r}'|^{-1}$, $\omega^\alpha_n$ is the plasmon frequency, $\epsilon_{avg}$ is the average dielectric constant of the media on both sides of the graphene layers, $\sigma$ is the DC conductivity of graphene, and $\tau$ is the Drude scattering time. The plasmon current density $\vec{K}$ of the coupled system satisfies the Drude equation,
\begin{equation}
\frac{\partial^{2} \vec{K}}{\partial t^{2}}  + \frac{1}{\tau}\frac{\partial \vec{K}}{\partial t} = \frac{\sigma}{\tau}\frac{\partial}{\partial t} \left( \vec{E}_{ext} + \vec{E}_{d}\right) \label{eq:K1}
\end{equation}
Here, $\vec{E}_{ext}$ is the external driving field and $\vec{E}_{d}$ is the depolarization field that results from the plasmon charge densities of all resonators,
\begin{equation} 
\frac{\partial  \vec{E}_{d}(\vec{r},t)}{\partial t} = \frac{-1}{4\pi\epsilon_{\textrm{avg}}}\int \textrm{d}^{2}\vec{r}' \bar{\bar{f}}(\vec{r}-\vec{r}') \cdot \vec{K}(\vec{r}',t) \label{eq:K2}
\end{equation}
We expand $\vec{K}$ in terms of the plasmon current density eigenmodes of different resonators, $\vec{K} = \sum_{\alpha,n} a^{\alpha}_{n}(t) \vec{K}^\alpha_n$. The plasmon eigenmodes are orthogonal in the sense,$\int d^{2}\vec{r} \vec{K}^n_{\alpha}(\vec{r})\cdot \vec{K}^m_{\beta}(\vec{r}) \, \tau/\sigma(\vec{r}) \propto \delta_{n m} \delta_{\alpha \beta}$, and if they are normalized such that the proportionality sign is an equality, then $|a^{\alpha}_{n}(t)|^{2}$ equals the energy in the plasmon eigenmode. Using the orthogonality of the plasmon eigenmodes, one obtains the coupled-mode equation,
\begin{eqnarray}
&& \frac{\partial^{2} a^{\alpha}_{n}(t)}{\partial t^{2}}  + \frac{1}{\tau}\frac{\partial a^{\alpha}_{n}(t)}{\partial t} + (\omega^\alpha_n)^{2} a^{\alpha}_{n}(t) = \nonumber \\
&& \omega^{\alpha}_n \omega^{\beta}_m \sum_{\substack{\beta\ne \alpha \\ m \ne n}} \Delta^{\alpha \beta}_{nm} a^{\beta}_{m}(t) +  \int d^{2}\vec{r} \vec{K}^\alpha_n(\vec{r}) \cdot \frac{\partial \vec{E}_{ext}(\vec{r},t)}{\partial t} 
\label{eq:K3}
\end{eqnarray}
The dimensionless and symmetric mode coupling parameters, $\Delta^{\alpha \beta}_{nm}$, can be expressed as,
\begin{equation} 
\Delta^{\alpha \beta}_{nm} = -\frac{C^{\alpha \beta}_{nm}}{\sqrt{C^{\alpha \alpha}_{nn} C^{\beta \beta}_{mm}}}
\label{eq:K4}
\end{equation}
where,
\begin{equation}
C^{\alpha \beta}_{nm} = \int \textrm{d}^{2}\vec{r} \int \textrm{d}^{2}\vec{r}' \vec{K}^{\alpha}_n(\vec{r}) \cdot \bar{\bar{f}}(\vec{r}-\vec{r}') \cdot \vec{K}^{\beta}_m(\vec{r}')
\end{equation}
The coupling parameters in Eq.\ref{eq:K4} determine the interactions between the plasmon modes and their symmetry with respect to the coupled modes is required for energy conservation. If the charge density associated with the plasmon mode is $\rho^{\alpha}_n(\vec{r}) \propto  \vec{\nabla}\cdot\vec{K}^{\alpha}_n(\vec{r})$, then the coupling parameter has a simple interpretation as the ratio of the cross-Coulomb energy to the self-Coulomb energy of the interacting plasmon modes. Although the form of Eq.\ref{eq:K3}, describing a set of coupled oscillators, is hardly surprising, it provides an accurate and systematic way to compute couplings between plasmon modes. Eq.\ref{eq:K3} shows that dark modes, for which the last term on the right hand side is zero, can get excited as a result of plasmon mode coupling~\cite{Shvets13,Shi13,Amin13}. Changes in the mode shape as a result of mode coupling~\cite{Strait13} is also captured in Eq.\ref{eq:K3} and results from the non-resonant excitation of higher order plasmon modes. We now present experimental results that verify the validity and accuracy of Equations \ref{eq:K3} and \ref{eq:K4}. 
\begin{figure}[tbp]
	\centering
		\includegraphics[width=.45\textwidth]{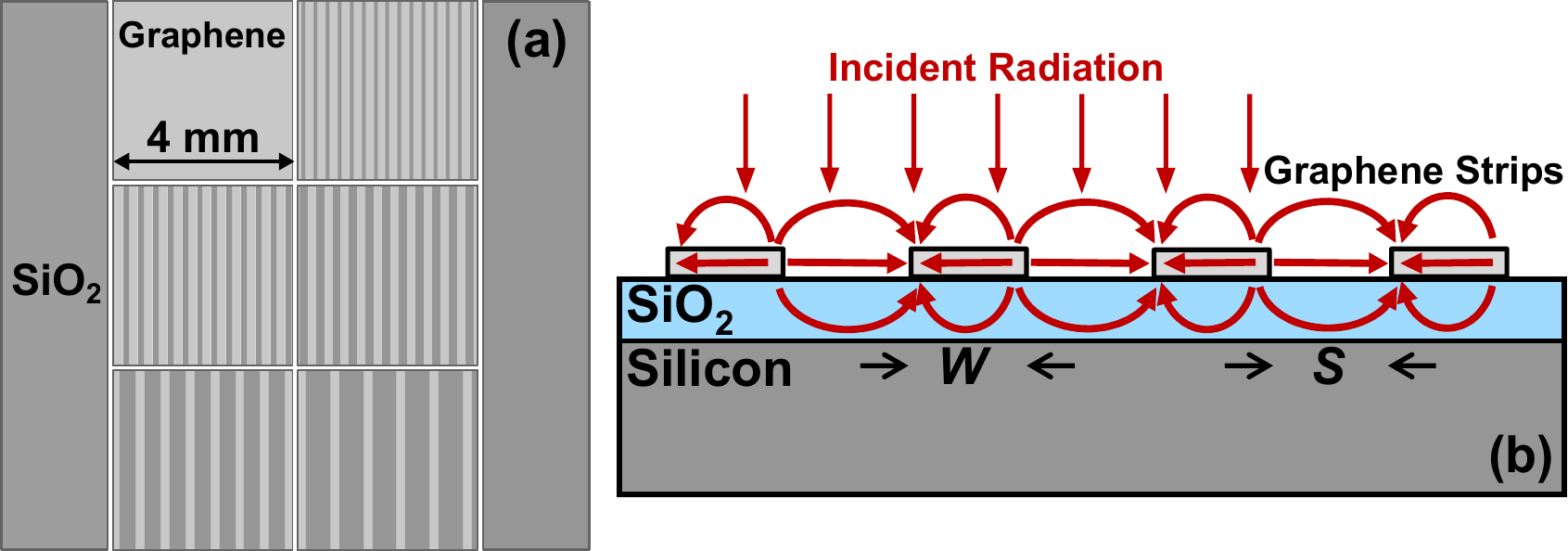}
	\caption{(a)Schematics of the sample (not to scale) showing six square regions of graphene strips with varying strip spacing and two silicon dioxide reference regions. The dark gray areas correspond to silicon dioxide and the light gray areas are graphene. (b) A cross section (not to scale) of an array of graphene strips with electric field lines corresponding to the lowest plasmon mode of the coupled system.}
	\label{fig:Schematics}
\end{figure}

\begin{figure}[tbp]
	\centering
		\includegraphics[width=.4\textwidth]{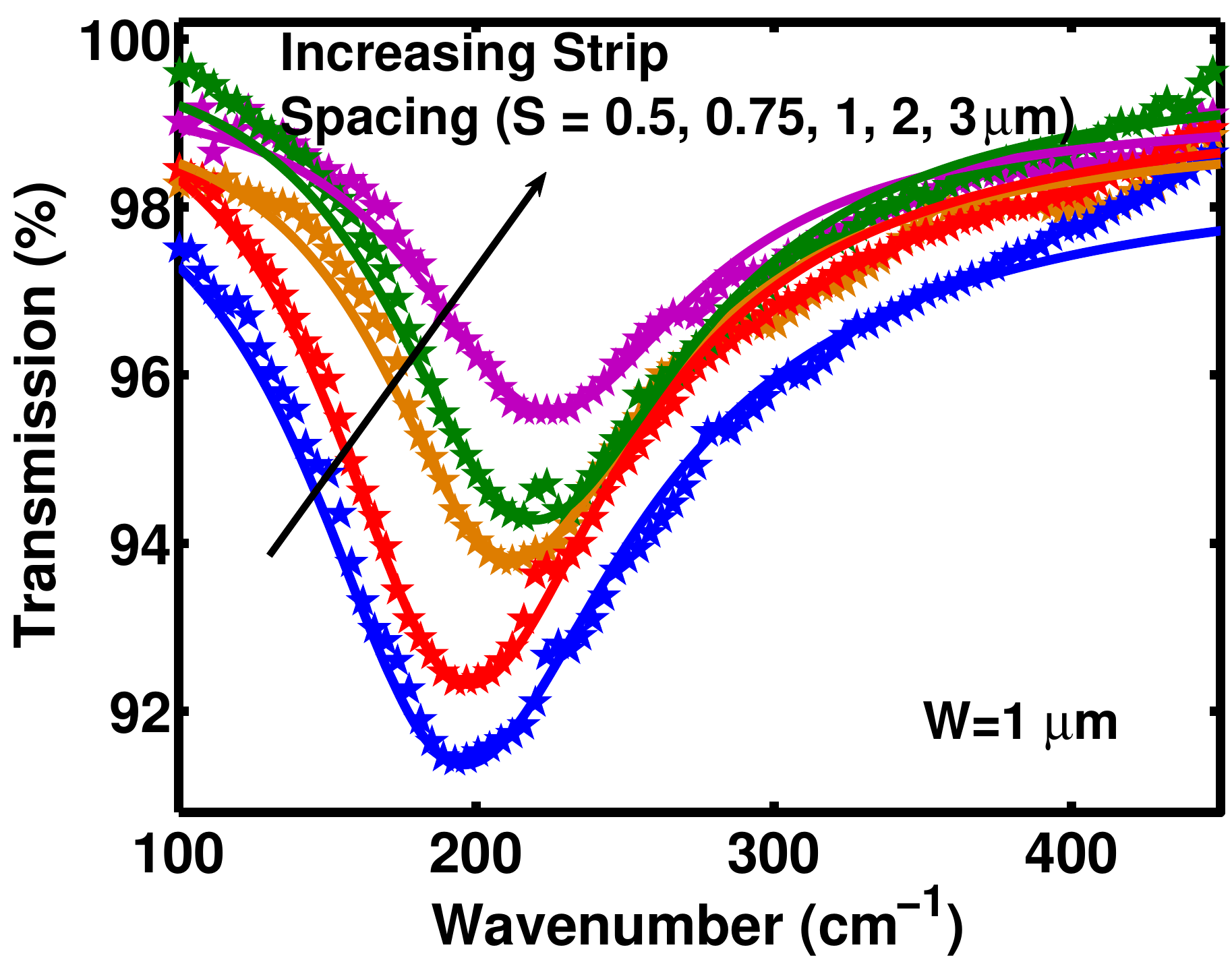}
	\caption{(Asterisks) Measured relative transmission for incident radiation polarized perpendicular to the graphene strip array (sample (b)). Strip width $W$ is 1 $\mu$m wide and strip spacings $S$ are 0.5, 0.75, 1, 2, and 3 $\mu$m. A bare SiO$_{2}$/Si substrate was used as reference. (Dashed lines) Fits to the measured data using a damped harmonic oscillator model in Eq.\ref{eq:Tr}.}
	\label{fig:PerpFits}
\end{figure}

Plasmonic structures studied in our experiments are shown in Fig.\ref{fig:Schematics} and consist of arrays of graphene strips of width $W$ = 1 $\mu$m and varying spacing, $S$~\cite{Ju11,HYan12,Strait13}. The spacing is varied to change the coupling strength among the plasmon modes of neighbouring strips and the resulting frequency shifts are measured experimentally. Two different graphene samples were used in the experiments; sample (a), grown by CVD on copper at Cornell, and sample (b), obtained from Graphenea and also grown by CVD on copper. Graphene was transferred onto high resistivity double-side polished silicon wafers (resistivity $>$ 10 $k\Omega$-cm) with $\sim$90 nm of thermally grown silicon dioxide~\cite{Li09}. Graphene strip arrays were patterned using lithography, etched using oxygen plasma, and then chemically doped by dipping the samples in nitric acid~\cite{Kasry10}. Measurements of plasmon resonance frequencies were carried out using a Fourier Transform Infrared Spectrometer (FTIR). Fig.\ref{fig:PerpFits} shows some typical transmission spectra of arrays with different strip spacings for incident light polarized perpendicular to the strips. A damped harmonic oscillator model was used to fit the transmission spectra and extract the plasmon frequency~\cite{Ju11,HYan12,Strait13},
\begin{equation}
\frac{T_{\textrm{sample}}(\omega)}{T_{\textrm{ref}}(\omega)} = \left| 1 + \frac{\eta_{o} f \sigma}{1 + n_{\textrm{sub}}}\frac{i\omega/\tau}{\omega^{2}-\omega_{p}^{2} + i\omega/\tau} \right|^{-2} \label{eq:Tr}
\end{equation}
Here, $\eta_{o}$ is the free-space impedance, $f$ is the fill factor of the strip array, $\sigma$ is the dc conductivity of graphene, $n_{\textrm{sub}}$ is the dielectric constant of the silicon substrate, $\tau$ is the plasmon damping time, and $\omega_{p}$ is the plasmon frequency. 
The transmission spectra show plasmon resonances~\cite{Ju11,HYan12,Strait13} whose frequencies decrease with the decrease in strip spacing. Although plasmon frequency redshifts as a result of interactions has been observed in graphene plasmon resonators previously~\cite{HYan12,Strait13}, the goal of this paper is to study  plasmon mode interactions in a quantitative way for the first time. We assume a uniform driving field of frequency $\omega$, $\vec{E}_{ext} = \vec{E}_{o}e^{-i\omega t}$, and expand the plasmon current density of the coupled system in terms of just the lowest plasmon modes ($\alpha=0$) of frequency $\omega^0$ of individual strips, $\vec{K} = \sum_{n} a^0_{n}e^{-i\omega t} \vec{K}^0_n$, and obtain the following coupled-mode equation,
\begin{eqnarray}
&& \left( -\omega^2 - i\frac{\omega}{\tau} + (\omega^0)^2 \right) a^0_{n} -(\omega^0)^2 \sum_{j=1,2}\left[ \Delta^{00}_{0j}(a^0_{n-j} + a^0_{n+j}) \right] \nonumber \\
&& = -i\omega \int d^{2}\vec{r} \vec{K}^0_n(\vec{r}) \cdot \vec{E}_{o} 
\label{eq:K5}
\end{eqnarray}
\begin{figure}[tbp]
	\centering
		\includegraphics[width=.4\textwidth]{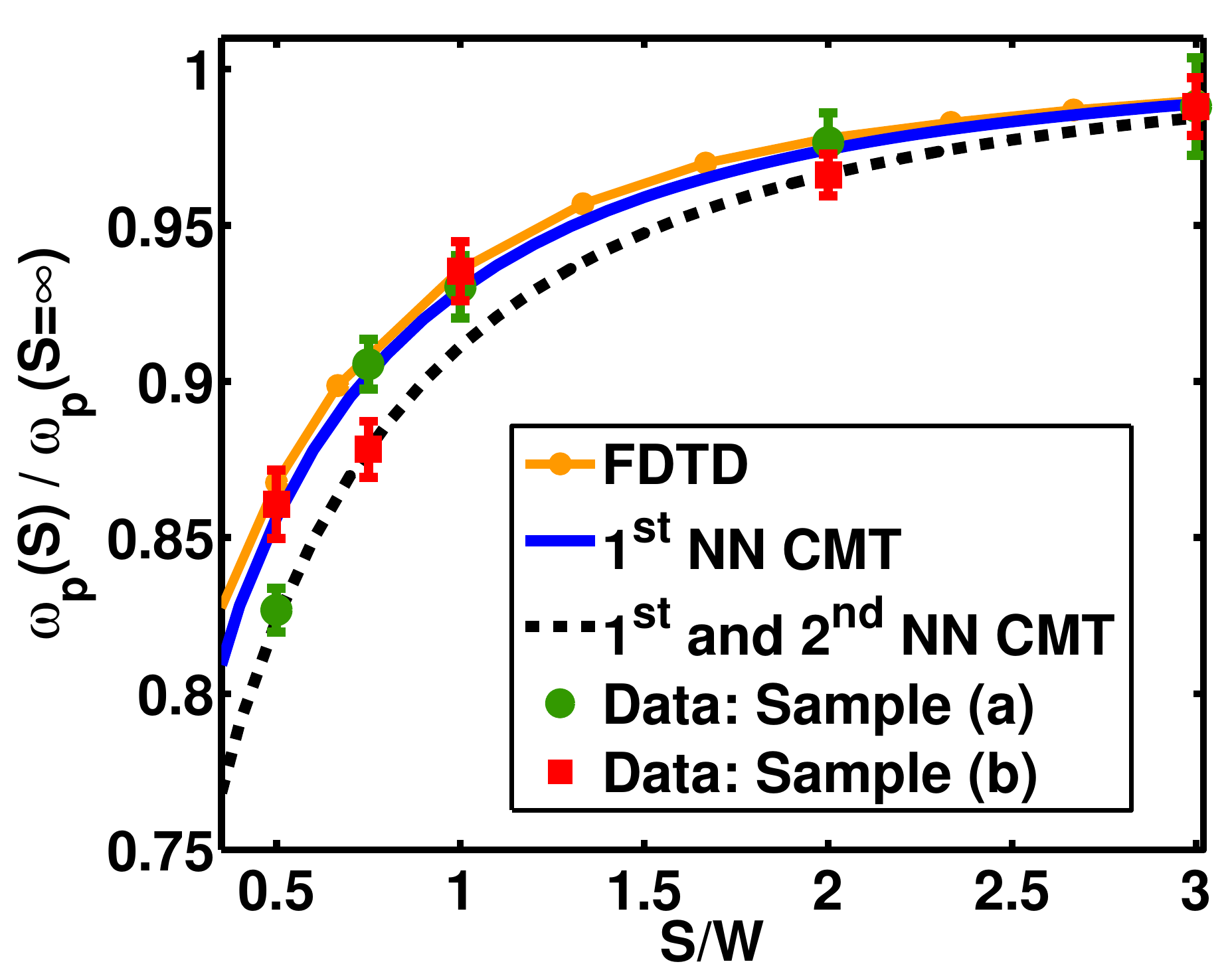}
	\caption{The measured (circles and squares) and calculated plasmon frequencies $\omega_{p}(S)$ of the lowest mode of a coupled strip array, normalized to a plasmon frequency of a single strip $\omega^0 = \omega_{p}(S=\infty)$, are plotted as a function of the strip-spacing to strip-width ratio ($S/W$). The calculations used coupled-mode theories (CMT) with only first nearest neighbour (NN) coupling (solid line), first and second nearest neighbour coupling (dashed line), and FDTD (solid line with circles).} 
	\label{fig:RedShift}
\end{figure}
$\Delta^{00}_{01}$ and $\Delta^{00}_{02}$ are the first and second nearest neighbour coupling parameters, respectively. A solution can be obtained for an infinite array by assuming that $a^0_{n} \propto e^{ikn(S+W)}$, and for the lowest plasmon mode of the coupled array ($k=0$) one obtains the resonance frequency $\omega_{p} = \omega^0\sqrt{1-2\Delta^{00}_{01}-2\Delta^{00}_{02}}$. Note that $\omega_{p}/\omega^0$ depends only on the geometrical ratio $S/W$. In calculations, first the lowest plasmon eigenmode of a single graphene strip was obtained by solving the eigenvalue equation (Eq.\ref{eq:K0}) using a 1D mesh with $1000$ points in the micron wide graphene strip and then the coupling parameters were computed using Eq.\ref{eq:K4}. FDTD was also used to obtain the plasmon frequencies of the coupled arrays using a non-uniform 2D mesh with $300$ points in the graphene strip and using an 10 element array (number of mesh points used was determined by the available computing power)~\cite{Strait13}. Fig.\ref{fig:RedShift} shows the measured and calculated plasmon frequency $\omega_{p}(S)$ of the lowest mode of a coupled strip array, normalized to a plasmon frequency of a single strip $\omega^0 = \omega_{p}(S=\infty)$, as a function of the ratio of the strip spacing to strip width ($S/W$). The effects of including only the first nearest neighbour coupling, and the first and second nearest neighbour couplings are also shown. The error bars indicate the accuracy with which the frequency could be measured in experiments given the signal to noise ratios. The error becomes large for spacings larger than 3 microns due to smaller fill factors. The remaining noise visible in the data (not represented by the error bars) is due to 10-15\% spatial variation in the doping in the samples which results in 2-3\% variation in the plasmon frequency. Within these error margins, the agreement between the coupled-mode theory and the experimental data appears to be very good. The reduction in the plasmon frequency of the coupled array observed in experiments due to interactions as the spacing between the strips is decreased is reproduced very accurately by the coupled-mode approach for all values of the ratio $S/W$ used in our experiments. Most measured data points fall within the margins defined by the first and second nearest neighbour coupling approximations. The slight discrepancy between the FDTD results and the coupled-mode theory results is attributd to two different facts: a) the 2D FDTD mesh used was not as fine as the 1D mesh used in the coupled-mode theory, and b) the FDTD used a 10 element array to simulate an infinite array because of limited computational resources. The mesh needs to be fine enough to resolve the plasmon charge density piled up at the edges of the graphene strips (for a strip centered at the origin with its width along the x-axis, the charge density varies with distance approximately as $x/\sqrt{(0.5W)^2 -x^2}$~\cite{Strait13}). For a finite $N$-element array, coupled-mode theory gives the plasmon frequency of the coupled array as, $\omega^0\sqrt{1-2\Delta^{00}_{01}\cos (\pi/N)}$ (assuming only nearest neighbour coupling) which shows that $N$ needs to be larger than 10 in order for the FDTD result to accurately match the result for an infinite array. 

Higher order corrections to the plasmon frequency of the coupled array resulting from interactions can also be computed using the coupled-mode theory. The second ($\alpha=1$) plasmonic mode of each strip is excited neither by the incident field nor by coupling among the strips because it does possess parity symmetry. Therefore, the next important correction comes from the non-resonant excitation of the third ($\alpha=2$) plasmonic mode of each strip from the depolarizing field of the nearest neighbours. This can also be understood as the change in the shape of the plasmon eigenmode of each strip as a result of interactions. So not surprisingly, the coupled-mode theory shows that the effect of this is to modify the value of the nearest neighbour coupling parameter $\Delta_{01}^{00}$ to,
\begin{equation}
\Delta_{01}^{00} +  2(\Delta_{01}^{02})^2 \frac{(\omega^2)^2}{(\omega^2)^2-(\omega^0)^2}
\end{equation}
The correction term, approximately $2.73(\Delta_{01}^{02})^2$, becomes larger for smaller spacings between the strips. For example, when $S/W=1$ the correction term is two orders of magnitude smaller than the second nearest neighbour coupling parameter $\Delta_{02}^{00}$. But when $S/W=0.1$ the correction term is of the same order of magnitude as $\Delta_{02}^{00}$. But for all values of $S/W$ in our devices, the correction term was found to be small and therefore ignored.  

To conclude, in this work we presented a coupled-mode theory for describing interactions between plasmon modes in graphene microstructures. The predictions of the coupled-mode theory as well as the accuracy with which it enabled the computation of the mode coupling parameters were tested in a quantitative way against experimental measurements. The agreement between the experiments and the theory was found to be very good. The work presented here provides a new tool to understand and design complex graphene plasmonic structures. 

The authors would like to acknowledge helpful discussions with Michael G. Spencer, help from Kevek Innovations Incorporated on graphene CVD growth, and support from CCMR under NSF grant number DMR-1120296, AFOSR-MURI under grant number FA9550-09-1-0705, ONR under grant number N00014-12-1-0072, and the Cornell Center for Nanoscale Systems funded by NSF. 

\bibliography{plasinter}

\begin{thebibliography}{24}
\expandafter\ifx\csname natexlab\endcsname\relax\def\natexlab#1{#1}\fi
\expandafter\ifx\csname bibnamefont\endcsname\relax
  \def\bibnamefont#1{#1}\fi
\expandafter\ifx\csname bibfnamefont\endcsname\relax
  \def\bibfnamefont#1{#1}\fi
\expandafter\ifx\csname citenamefont\endcsname\relax
  \def\citenamefont#1{#1}\fi
\expandafter\ifx\csname url\endcsname\relax
  \def\url#1{\texttt{#1}}\fi
\expandafter\ifx\csname urlprefix\endcsname\relax\def\urlprefix{URL }\fi
\providecommand{\bibinfo}[2]{#2}
\providecommand{\eprint}[2][]{\url{#2}}

\bibitem[{\citenamefont{Ju et~al.}(2011)\citenamefont{Ju, Geng, Horng, Girit,
  Martin, Hao, Bechtel, Zettl, Shen, and Wang}}]{Ju11}
\bibinfo{author}{\bibfnamefont{L.}~\bibnamefont{Ju}},
  \bibinfo{author}{\bibfnamefont{B.}~\bibnamefont{Geng}},
  \bibinfo{author}{\bibfnamefont{J.}~\bibnamefont{Horng}},
  \bibinfo{author}{\bibfnamefont{C.}~\bibnamefont{Girit}},
  \bibinfo{author}{\bibfnamefont{M.}~\bibnamefont{Martin}},
  \bibinfo{author}{\bibfnamefont{Z.}~\bibnamefont{Hao}},
  \bibinfo{author}{\bibfnamefont{H.~A.} \bibnamefont{Bechtel}},
  \bibinfo{author}{\bibfnamefont{A.}~\bibnamefont{Zettl}},
  \bibinfo{author}{\bibfnamefont{Y.~R.} \bibnamefont{Shen}}, \bibnamefont{and}
  \bibinfo{author}{\bibfnamefont{F.}~\bibnamefont{Wang}},
  \bibinfo{journal}{Nature Nanotechnology} \textbf{\bibinfo{volume}{6}},
  \bibinfo{pages}{630} (\bibinfo{year}{2011}).

\bibitem[{\citenamefont{Yan et~al.}(2012)\citenamefont{Yan, Li, Chandra,
  Tulevski, Wu, Freitag, Zhu, Avouris, and Xia}}]{HYan12}
\bibinfo{author}{\bibfnamefont{H.}~\bibnamefont{Yan}},
  \bibinfo{author}{\bibfnamefont{X.}~\bibnamefont{Li}},
  \bibinfo{author}{\bibfnamefont{B.}~\bibnamefont{Chandra}},
  \bibinfo{author}{\bibfnamefont{G.}~\bibnamefont{Tulevski}},
  \bibinfo{author}{\bibfnamefont{Y.}~\bibnamefont{Wu}},
  \bibinfo{author}{\bibfnamefont{M.}~\bibnamefont{Freitag}},
  \bibinfo{author}{\bibfnamefont{W.}~\bibnamefont{Zhu}},
  \bibinfo{author}{\bibfnamefont{P.}~\bibnamefont{Avouris}}, \bibnamefont{and}
  \bibinfo{author}{\bibfnamefont{F.}~\bibnamefont{Xia}},
  \bibinfo{journal}{Nature Nanotechnology} \textbf{\bibinfo{volume}{7}},
  \bibinfo{pages}{330} (\bibinfo{year}{2012}).

\bibitem[{\citenamefont{Rana}(2008)}]{Rana08}
\bibinfo{author}{\bibfnamefont{F.}~\bibnamefont{Rana}}, \bibinfo{journal}{IEEE
  Trans. Nano.} \textbf{\bibinfo{volume}{7}}, \bibinfo{pages}{91}
  (\bibinfo{year}{2008}).

\bibitem[{\citenamefont{Das~Sarma and Hwang}(2009)}]{DasSarma09}
\bibinfo{author}{\bibfnamefont{S.}~\bibnamefont{Das~Sarma}} \bibnamefont{and}
  \bibinfo{author}{\bibfnamefont{E.~H.} \bibnamefont{Hwang}},
  \bibinfo{journal}{Phys. Rev. Lett.} \textbf{\bibinfo{volume}{102}},
  \bibinfo{pages}{206412} (\bibinfo{year}{2009}).

\bibitem[{\citenamefont{Hwang and Das~Sarma}(2007)}]{Hwang07}
\bibinfo{author}{\bibfnamefont{E.~H.} \bibnamefont{Hwang}} \bibnamefont{and}
  \bibinfo{author}{\bibfnamefont{S.}~\bibnamefont{Das~Sarma}},
  \bibinfo{journal}{Phys. Rev. B} \textbf{\bibinfo{volume}{75}},
  \bibinfo{pages}{205418} (\bibinfo{year}{2007}).

\bibitem[{\citenamefont{Yan et~al.}(2013)\citenamefont{Yan, Low, Zhu, Wu,
  Freitag, Li, Guinea, Avouris, and Xia}}]{Xia12}
\bibinfo{author}{\bibfnamefont{H.}~\bibnamefont{Yan}},
  \bibinfo{author}{\bibfnamefont{T.}~\bibnamefont{Low}},
  \bibinfo{author}{\bibfnamefont{W.}~\bibnamefont{Zhu}},
  \bibinfo{author}{\bibfnamefont{Y.}~\bibnamefont{Wu}},
  \bibinfo{author}{\bibfnamefont{M.}~\bibnamefont{Freitag}},
  \bibinfo{author}{\bibfnamefont{X.}~\bibnamefont{Li}},
  \bibinfo{author}{\bibfnamefont{F.}~\bibnamefont{Guinea}},
  \bibinfo{author}{\bibfnamefont{P.}~\bibnamefont{Avouris}}, \bibnamefont{and}
  \bibinfo{author}{\bibfnamefont{F.}~\bibnamefont{Xia}},
  \bibinfo{journal}{Nature Photonics} \textbf{\bibinfo{volume}{7}},
  \bibinfo{pages}{394} (\bibinfo{year}{2013}).

\bibitem[{\citenamefont{Koppens et~al.}(2011)\citenamefont{Koppens, Chang, and
  García~de Abajo}}]{Koppens11}
\bibinfo{author}{\bibfnamefont{F.~H.~L.} \bibnamefont{Koppens}},
  \bibinfo{author}{\bibfnamefont{D.~E.} \bibnamefont{Chang}}, \bibnamefont{and}
  \bibinfo{author}{\bibfnamefont{F.~J.} \bibnamefont{García~de Abajo}},
  \bibinfo{journal}{Nano Letters} \textbf{\bibinfo{volume}{11}},
  \bibinfo{pages}{3370} (\bibinfo{year}{2011}).

\bibitem[{\citenamefont{Low and Avouris}(2014)}]{Low14}
\bibinfo{author}{\bibfnamefont{T.}~\bibnamefont{Low}} \bibnamefont{and}
  \bibinfo{author}{\bibfnamefont{P.}~\bibnamefont{Avouris}},
  \bibinfo{journal}{ACS Nano} \textbf{\bibinfo{volume}{8}},
  \bibinfo{pages}{1086} (\bibinfo{year}{2014}).

\bibitem[{\citenamefont{Peres et~al.}(2013)\citenamefont{Peres, Bludov,
  Ferreira, and Vasilevskiy}}]{Peres13}
\bibinfo{author}{\bibfnamefont{N.~M.~R.} \bibnamefont{Peres}},
  \bibinfo{author}{\bibfnamefont{Y.~V.} \bibnamefont{Bludov}},
  \bibinfo{author}{\bibfnamefont{A.}~\bibnamefont{Ferreira}}, \bibnamefont{and}
  \bibinfo{author}{\bibfnamefont{M.~I.} \bibnamefont{Vasilevskiy}},
  \bibinfo{journal}{Journal of Physics: Condensed Matter}
  \textbf{\bibinfo{volume}{25}}, \bibinfo{pages}{125303}
  (\bibinfo{year}{2013}).

\bibitem[{\citenamefont{Fallahi and Perruisseau-Carrier}(2012)}]{Fallahi12}
\bibinfo{author}{\bibfnamefont{A.}~\bibnamefont{Fallahi}} \bibnamefont{and}
  \bibinfo{author}{\bibfnamefont{J.}~\bibnamefont{Perruisseau-Carrier}},
  \bibinfo{journal}{Phys. Rev. B} \textbf{\bibinfo{volume}{86}},
  \bibinfo{pages}{195408} (\bibinfo{year}{2012}).

\bibitem[{\citenamefont{Thongrattanasiri
  et~al.}(2012)\citenamefont{Thongrattanasiri, Koppens, and Garc\'\! ia~de
  Abajo}}]{Thongrattanasiri12}
\bibinfo{author}{\bibfnamefont{S.}~\bibnamefont{Thongrattanasiri}},
  \bibinfo{author}{\bibfnamefont{F.~H.~L.} \bibnamefont{Koppens}},
  \bibnamefont{and} \bibinfo{author}{\bibfnamefont{F.~J.} \bibnamefont{Garc\'\!
  ia~de Abajo}}, \bibinfo{journal}{Phys. Rev. Lett.}
  \textbf{\bibinfo{volume}{108}}, \bibinfo{pages}{047401}
  (\bibinfo{year}{2012}).

\bibitem[{\citenamefont{Hanson}(2008)}]{Hanson08}
\bibinfo{author}{\bibfnamefont{G.~H.} \bibnamefont{Hanson}},
  \bibinfo{journal}{J. App. Phys.} \textbf{\bibinfo{volume}{103}},
  \bibinfo{pages}{064302} (\bibinfo{year}{2008}).

\bibitem[{\citenamefont{Ferreira and Peres}(2012)}]{Ferreira12}
\bibinfo{author}{\bibfnamefont{A.}~\bibnamefont{Ferreira}} \bibnamefont{and}
  \bibinfo{author}{\bibfnamefont{N.~M.~R.} \bibnamefont{Peres}},
  \bibinfo{journal}{Phys. Rev. B} \textbf{\bibinfo{volume}{86}},
  \bibinfo{pages}{205401} (\bibinfo{year}{2012}).

\bibitem[{\citenamefont{Strait et~al.}(2013)\citenamefont{Strait, Nene, Chan,
  Manolatou, Tiwari, Rana, Kevek, and McEuen}}]{Strait13}
\bibinfo{author}{\bibfnamefont{J.~H.} \bibnamefont{Strait}},
  \bibinfo{author}{\bibfnamefont{P.}~\bibnamefont{Nene}},
  \bibinfo{author}{\bibfnamefont{W.-M.} \bibnamefont{Chan}},
  \bibinfo{author}{\bibfnamefont{C.}~\bibnamefont{Manolatou}},
  \bibinfo{author}{\bibfnamefont{S.}~\bibnamefont{Tiwari}},
  \bibinfo{author}{\bibfnamefont{F.}~\bibnamefont{Rana}},
  \bibinfo{author}{\bibfnamefont{J.~W.} \bibnamefont{Kevek}}, \bibnamefont{and}
  \bibinfo{author}{\bibfnamefont{P.~L.} \bibnamefont{McEuen}},
  \bibinfo{journal}{Phys. Rev. B} \textbf{\bibinfo{volume}{87}},
  \bibinfo{pages}{241410} (\bibinfo{year}{2013}).

\bibitem[{\citenamefont{Haus and Huang}(1991)}]{Haus91}
\bibinfo{author}{\bibfnamefont{H.}~\bibnamefont{Haus}} \bibnamefont{and}
  \bibinfo{author}{\bibfnamefont{W.-P.} \bibnamefont{Huang}},
  \bibinfo{journal}{Proceedings of the IEEE} \textbf{\bibinfo{volume}{79}},
  \bibinfo{pages}{1505} (\bibinfo{year}{1991}), ISSN \bibinfo{issn}{0018-9219}.

\bibitem[{\citenamefont{Yariv}(1973)}]{Yariv73}
\bibinfo{author}{\bibfnamefont{A.}~\bibnamefont{Yariv}},
  \bibinfo{journal}{Quantum Electronics, IEEE Journal of}
  \textbf{\bibinfo{volume}{9}}, \bibinfo{pages}{919} (\bibinfo{year}{1973}),
  ISSN \bibinfo{issn}{0018-9197}.

\bibitem[{\citenamefont{Manolatou and Haus}(2001)}]{Manolatou01}
\bibinfo{author}{\bibfnamefont{C.}~\bibnamefont{Manolatou}} \bibnamefont{and}
  \bibinfo{author}{\bibfnamefont{H.~A.} \bibnamefont{Haus}},
  \emph{\bibinfo{title}{Passive Components for Dense Optical Integration}}
  (\bibinfo{publisher}{Springer}, \bibinfo{year}{2001}), \bibinfo{edition}{1st}
  ed., ISBN \bibinfo{isbn}{146135272X}.

\bibitem[{\citenamefont{Mousavi et~al.}(2013)\citenamefont{Mousavi, Kholmanov,
  Alici, Purtseladze, Arju, Tatar, Fozdar, Suk, Hao, Khanikaev
  et~al.}}]{Shvets13}
\bibinfo{author}{\bibfnamefont{S.~H.} \bibnamefont{Mousavi}},
  \bibinfo{author}{\bibfnamefont{I.}~\bibnamefont{Kholmanov}},
  \bibinfo{author}{\bibfnamefont{K.~B.} \bibnamefont{Alici}},
  \bibinfo{author}{\bibfnamefont{D.}~\bibnamefont{Purtseladze}},
  \bibinfo{author}{\bibfnamefont{N.}~\bibnamefont{Arju}},
  \bibinfo{author}{\bibfnamefont{K.}~\bibnamefont{Tatar}},
  \bibinfo{author}{\bibfnamefont{D.~Y.} \bibnamefont{Fozdar}},
  \bibinfo{author}{\bibfnamefont{J.~W.} \bibnamefont{Suk}},
  \bibinfo{author}{\bibfnamefont{Y.}~\bibnamefont{Hao}},
  \bibinfo{author}{\bibfnamefont{A.~B.} \bibnamefont{Khanikaev}},
  \bibnamefont{et~al.}, \bibinfo{journal}{Nano Letters}
  \textbf{\bibinfo{volume}{13}}, \bibinfo{pages}{1111} (\bibinfo{year}{2013}).

\bibitem[{\citenamefont{Shi et~al.}(2013)\citenamefont{Shi, Han, Dai, Yu, Sun,
  Chen, Liu, and Zi}}]{Shi13}
\bibinfo{author}{\bibfnamefont{X.}~\bibnamefont{Shi}},
  \bibinfo{author}{\bibfnamefont{D.}~\bibnamefont{Han}},
  \bibinfo{author}{\bibfnamefont{Y.}~\bibnamefont{Dai}},
  \bibinfo{author}{\bibfnamefont{Z.}~\bibnamefont{Yu}},
  \bibinfo{author}{\bibfnamefont{Y.}~\bibnamefont{Sun}},
  \bibinfo{author}{\bibfnamefont{H.}~\bibnamefont{Chen}},
  \bibinfo{author}{\bibfnamefont{X.}~\bibnamefont{Liu}}, \bibnamefont{and}
  \bibinfo{author}{\bibfnamefont{J.}~\bibnamefont{Zi}}, \bibinfo{journal}{Opt.
  Express} \textbf{\bibinfo{volume}{21}}, \bibinfo{pages}{28438}
  (\bibinfo{year}{2013}).

\bibitem[{\citenamefont{Amin et~al.}(2013)\citenamefont{Amin, Farhat, and
  Bagci}}]{Amin13}
\bibinfo{author}{\bibfnamefont{M.}~\bibnamefont{Amin}},
  \bibinfo{author}{\bibfnamefont{M.}~\bibnamefont{Farhat}}, \bibnamefont{and}
  \bibinfo{author}{\bibfnamefont{H.}~\bibnamefont{Bagci}},
  \bibinfo{journal}{Sci. Rep.} \textbf{\bibinfo{volume}{3}}, \bibinfo{pages}{8}
  (\bibinfo{year}{2013}).

\bibitem[{\citenamefont{Fang et~al.}(2014)\citenamefont{Fang, Wang, Schlather,
  Liu, Ajayan, GarcÃ­a~de Abajo, Nordlander, Zhu, and Halas}}]{Fang14}
\bibinfo{author}{\bibfnamefont{Z.}~\bibnamefont{Fang}},
  \bibinfo{author}{\bibfnamefont{Y.}~\bibnamefont{Wang}},
  \bibinfo{author}{\bibfnamefont{A.~E.} \bibnamefont{Schlather}},
  \bibinfo{author}{\bibfnamefont{Z.}~\bibnamefont{Liu}},
  \bibinfo{author}{\bibfnamefont{P.~M.} \bibnamefont{Ajayan}},
  \bibinfo{author}{\bibfnamefont{F.~J.} \bibnamefont{GarcÃ­a~de Abajo}},
  \bibinfo{author}{\bibfnamefont{P.}~\bibnamefont{Nordlander}},
  \bibinfo{author}{\bibfnamefont{X.}~\bibnamefont{Zhu}}, \bibnamefont{and}
  \bibinfo{author}{\bibfnamefont{N.~J.} \bibnamefont{Halas}},
  \bibinfo{journal}{Nano Letters} \textbf{\bibinfo{volume}{14}},
  \bibinfo{pages}{299} (\bibinfo{year}{2014}).

\bibitem[{\citenamefont{Sensale-Rodriguez
  et~al.}(2012)\citenamefont{Sensale-Rodriguez, Yan, Zhu, Jena, Liu, and
  Grace~Xing}}]{Sensale12}
\bibinfo{author}{\bibfnamefont{B.}~\bibnamefont{Sensale-Rodriguez}},
  \bibinfo{author}{\bibfnamefont{R.}~\bibnamefont{Yan}},
  \bibinfo{author}{\bibfnamefont{M.}~\bibnamefont{Zhu}},
  \bibinfo{author}{\bibfnamefont{D.}~\bibnamefont{Jena}},
  \bibinfo{author}{\bibfnamefont{L.}~\bibnamefont{Liu}}, \bibnamefont{and}
  \bibinfo{author}{\bibfnamefont{H.}~\bibnamefont{Grace~Xing}},
  \bibinfo{journal}{Applied Physics Letters} \textbf{\bibinfo{volume}{101}},
  \bibinfo{eid}{261115} (\bibinfo{year}{2012}).

\bibitem[{\citenamefont{Li et~al.}(2009)\citenamefont{Li, Cai, An, Kim, Nah,
  Yang, Piner, Velamakanni, Jung, Tutuc et~al.}}]{Li09}
\bibinfo{author}{\bibfnamefont{X.}~\bibnamefont{Li}},
  \bibinfo{author}{\bibfnamefont{W.}~\bibnamefont{Cai}},
  \bibinfo{author}{\bibfnamefont{J.}~\bibnamefont{An}},
  \bibinfo{author}{\bibfnamefont{S.}~\bibnamefont{Kim}},
  \bibinfo{author}{\bibfnamefont{J.}~\bibnamefont{Nah}},
  \bibinfo{author}{\bibfnamefont{D.}~\bibnamefont{Yang}},
  \bibinfo{author}{\bibfnamefont{R.}~\bibnamefont{Piner}},
  \bibinfo{author}{\bibfnamefont{A.}~\bibnamefont{Velamakanni}},
  \bibinfo{author}{\bibfnamefont{I.}~\bibnamefont{Jung}},
  \bibinfo{author}{\bibfnamefont{E.}~\bibnamefont{Tutuc}},
  \bibnamefont{et~al.}, \bibinfo{journal}{Science}
  \textbf{\bibinfo{volume}{324}}, \bibinfo{pages}{1312} (\bibinfo{year}{2009}).

\bibitem[{\citenamefont{Kasry et~al.}(2010)\citenamefont{Kasry, Kuroda,
  Martyna, Tulevski, and Bol}}]{Kasry10}
\bibinfo{author}{\bibfnamefont{A.}~\bibnamefont{Kasry}},
  \bibinfo{author}{\bibfnamefont{M.~A.} \bibnamefont{Kuroda}},
  \bibinfo{author}{\bibfnamefont{G.~J.} \bibnamefont{Martyna}},
  \bibinfo{author}{\bibfnamefont{G.~S.} \bibnamefont{Tulevski}},
  \bibnamefont{and} \bibinfo{author}{\bibfnamefont{A.~A.} \bibnamefont{Bol}},
  \bibinfo{journal}{ACS Nano} \textbf{\bibinfo{volume}{4}},
  \bibinfo{pages}{3839} (\bibinfo{year}{2010}).

\end{thebibliography}

\end{document}